# MatchRDMA: A Segmented and Rate-Matched Long-Haul RDMA Scheme for Geo-distributed LLM Training over OTN

Jun Dai[(1)], Xiaorun Wang[(1)], Xingde Li[(1)], Zheng Yang[(1)], Kexiong Fang[(1)], Zhiqun Gu[(1)], Hongxiang Wang[(1)], Yuefeng Ji[(1)], and Jiawei Zhang[(1)*]

[(1)] State Key Lab of Information Photonics and Optical Communications, Beijing University of Posts and Telecommunications (BUPT), Beijing, 100876, China. *Corresponding author email: zjw@bupt.edu.cn

**Abstract** *We propose MatchRDMA, a proactive, segmented, and rate-matched long-haul RDMA scheme for geo-distributed LLM training over OTN. By coordinating source and destination OTN rates, it improves inter-DC throughput by up to 20× compared with conventional RDMA, and reduces destination-OTN buffer occupancy by up to 62.7%.* ©2026 The Author(s)

## Introduction

Remote Direct Memory Access (RDMA) has been widely adopted in datacenter and high-performance computing environments due to its low latency, high bandwidth efficiency, and minimal CPU overhead [1-4], and is increasingly important for large language model (LLM) training [5]. Meanwhile, the rapid growth of model parameters is pushing training demand beyond the capacity of a single AI datacenter (AI-DC), making geo-distributed training across multiple AI-DCs increasingly necessary [6-9]. However, extending conventional RDMA over Optical Transport Networks (OTNs) for geo-distributed training introduces significant performance bottlenecks due to the long-haul transmission. As shown in Fig. 1, three bottlenecks emerge: (1) **ACK-limited progress**. RDMA relies on acknowledgments (ACKs) returned by the receiver to confirm successful reception of earlier data (e.g., RoCE packets), thereby allowing the sender to continue transmitting new data. Over long-haul transmission, this ACK-return loop is significantly prolonged, slowing sender progress and reducing effective throughput. (2) **Buffer stress**. A longer inter-DC path means that more data remains in flight along the end-to-end transport path. If downstream forwarding is temporarily slowed by a congestion (as shown in Fig. 1), this excess in-flight data imposes strong buffer pressure near the destination OTN node; if the available buffer is insufficient, packet loss and inefficient retransmission become more likely. (3) **Feedback imbalance**. End-to-end congestion-control schemes designed for short and relatively uniform intra-DC feedback loops become much less effective in inter-DC settings. When inter-DC and intra-DC traffic with different feedback delays jointly contribute to a congestion, rate adaptation slows and bandwidth sharing becomes increasingly unbalanced. Together, these effects reduce transmission efficiency, exacerbate buffer stress at the destination OTN, and complicate efficient coordination between inter-DC and intra-DC traffic.

Existing studies have improved long-haul

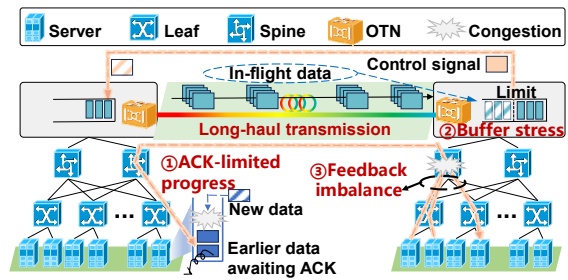

**Fig. 1:** Three bottlenecks for long-haul RDMA over OTN.

RDMA by accelerating ACK return [10], introducing relay-assisted control [11,12], and mitigating unfairness caused by different feedback delays [13-15]. Overall, these approaches are mainly reactive, as they respond only after congestion has already emerged, and are largely designed for the small, irregular, and hard-to-predict traffic patterns that characterize conventional DC workloads.

In this paper, we propose MatchRDMA, a proactive, segmented, and rate-matched long-haul RDMA scheme for geo-distributed LLM training over OTN. Leveraging the predictable communication structure of LLM training [16,17], MatchRDMA coordinates inter-DC transmission through segmented control and source-to-destination rate matching. It is realized via source-OTN budget-gated pseudo-ACK generation and congestion-control proxying, inter-OTN rate-budget signaling, and destination-OTN communication-aware slot-weighted rate estimation. The simulation results demonstrate that, compared with conventional RDMA, MatchRDMA improves inter-DC throughput by up to 20×, reduces destination-OTN buffer occupancy by up to 62.7%.

## Principles of MatchRDMA

Fig. 2(a) compares the workflow of conventional end-to-end RDMA with the segmented OTN-assisted MatchRDMA. Unlike conventional RDMA over long-haul, where the OTN acts only as a passive transport pipe, MatchRDMA turns the source and destination OTN nodes into active control elements and decomposes the round-trip time (RTT)-driven control path into three coordinated segments: a source-side loop, an inter-OTN coordination loop, and a destination-side loop (as shown in Fig. 2(a)). In lossless RDMA transport, sender progress depends on ACK return, while congestion control depends on explicit congestion notification (ECN) marking and congestion notification packet (CNP) feedback. Over a long-haul OTN transmission, both the ACK-return loop and the ECN/CNP-based control loop are significantly stretched by the long RTT. By splitting this long RTT-driven control path into three segments, MatchRDMA makes sender advancement and congestion regulation more responsive, thereby mitigating ACK-limited progress and feedback imbalance.

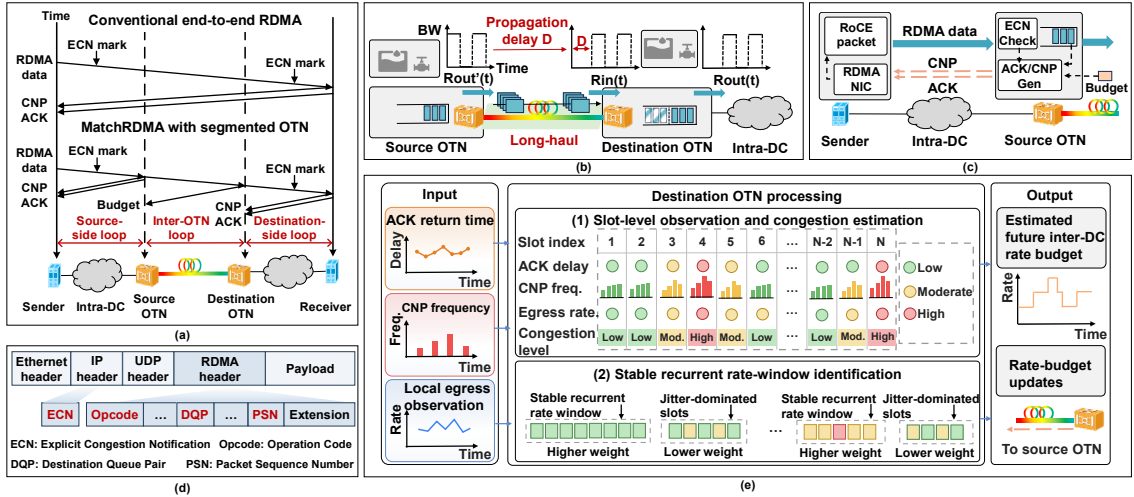

**Fig. 2:** Principles of MatchRDMA. (a) Comparison of conventional long-haul RDMA and MatchRDMA with segmented OTN-assisted control; (b) Reservoir model of destination-OTN buffer stress; (c) Source-OTN control workflow; (d) RoCE packet fields used for OTN-side control; (e) Destination-OTN slot-level rate estimation and rate-budget generation.

While segmented control alleviates two bottlenecks, the long-haul distance still induces destination OTN buffer stress. As illustrated in Fig. 2(b), we model the source and destination OTN nodes as two coupled reservoirs linked by a long-haul optical pipe. Let $D$ denote the one-way propagation delay. The arrival process at the destination OTN is then simply the delayed version of the source-OTN output process. If the source injects traffic at a time-varying rate $r_{in}(t)$ while the destination can forward traffic into the receiving AI-DC only at rate $r_{out}(t)$, the minimum runtime buffer required at the destination is governed by the accumulated rate mismatch over the control-uncertainty window. This can be expressed as:

$$B_{req} \geq \sup_{t} \int_{t}^{t+\tau} (r_{in}(u) - r_{out}(u))^{+} du \quad (1)$$

where $\tau$ denotes the effective time window during which source injection cannot be instantly adjusted—due to propagation and processing delays. Since LLM training traffic exhibits predictable communication structure, $r_{in}(t)$ can be proactively shaped to follow the expected destination-side rate, directly motivating the proactive, rate-matched principle of MatchRDMA.

Guided by these principles, MatchRDMA realizes proactive, segmented, and rate-matched control by inserting a lightweight RDMA-aware layer at the OTN nodes. As illustrated in Fig. 2(c), the source OTN learns RDMA connection-state information during flow setup and, during transmission, parses ECN marks together with selected RDMA header fields; Fig. 2(d) shows the RDMA-over-Converged Ethernet (RoCE) packet fields used for connection identification and sequence tracking. Based on this RDMA-aware view, the source OTN generates a budget-gated pseudo-ACK and a source-side congestion-control signal. The budget refers to the maximum sustainable injection rate that the destination side can accommodate. This accelerates sender progress while constraining inter-DC injection within the destination-sustainable rate budget, thereby preserving destination-side rate matching and mitigating unfairness caused by the interaction with intra-DC feedback loops.

Across the inter-OTN path, the two OTN nodes measure the one-way propagation delay and exchange rate-budget updates, along with concise congestion summaries, over a small high-priority control subchannel. This ensures that source-side release remains matched to the forwarding capability estimated at the destination side.

At the destination OTN, MatchRDMA uses returned CNPs for reactive rate tightening, but more importantly constructs communication-aware slot-level observations from ACK/CNP feedback and local egress observations. As illustrated in Fig. 2(e), ACK return time and CNP generation frequency are compared against preset thresholds to assess the congestion level in each slot. By aggregating consecutive slots, the destination OTN identifies stable recurrent rate windows and distinguishes them from short-term jitter-dominated slots. This communication-aware view is then used to estimate the future inter-DC rate budget and generate the corresponding rate-budget updates fed back to the source OTN. For robustness, unstable short-term slots are weighted conservatively, whereas stable recurrent rate windows receive higher weights. In this way, MatchRDMA proactively coordinates source admission with destination-side rate estimation under segmented control.

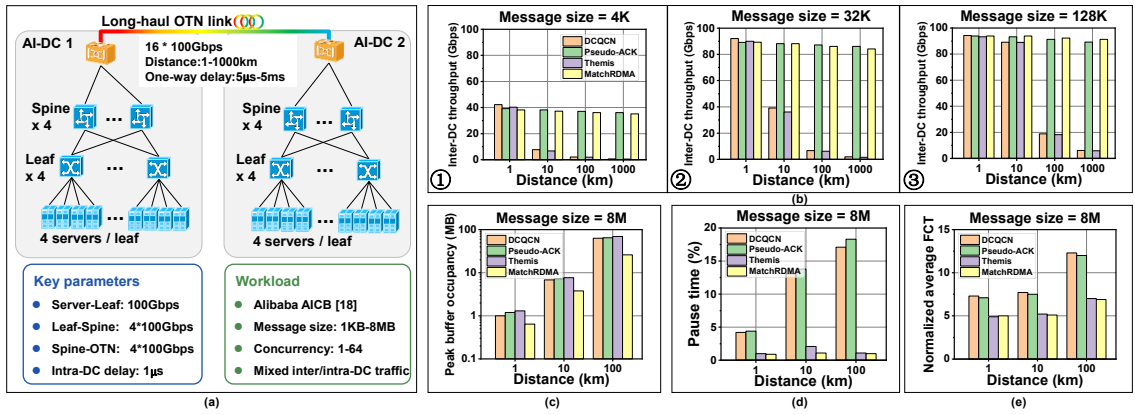

**Fig. 3:** (a) Simulated dual-AI-DC leaf-spine-OTN topology; (b) Throughput vs. distance under different message size; (c) Destination OTN runtime buffer occupancy; (d) Pause time ratio; (e) Overall average FCT in the mixed-traffic scenario.

**Simulation setup and results**
We evaluate MatchRDMA using the ns-3 simulator. The simulated topology, shown in Fig. 3(a), consists of two AI-DCs interconnected by 16 bidirectional OTN links, each operating at 100 Gbps. The one-way intra-DC delay is fixed at $1\ \mu s$, while the inter-DC distance is varied from $1\ km$ to $1000\ km$, corresponding to a one-way propagation delay from $5\ \mu s$ to $5\ ms$. The AI training workload is generated by the Alibaba AICB benchmark [18], which captures the alternating computation–communication structure of LLM training iterations. In the simulator, each training communication request is modeled as a message, which may be segmented into one or more RoCE packets depending on its size. We sweep the message size from 1 KB to 8 MB and the number of parallel messages (concurrency) from 1 to 64. MatchRDMA is compared with three baselines: a conventional DCQCN-like baseline [1], a pseudo-ACK baseline by NTT [10], and a THEMIS-like fairness-oriented baseline [14,19].

Fig. 3(b) shows the inter-DC throughput under different message sizes and inter-DC OTN delays. As the delay increases, the DCQCN-like and THEMIS-like baselines suffer severe throughput degradation because sender progress remains constrained by the stretched end-to-end ACK feedback loop. In contrast, both pseudo-ACK baseline and MatchRDMA are much less sensitive to distance, since pseudo-ACK keeps sender progress active. In addition, throughput increases with message size for all schemes because larger messages make better use of the long-haul path. Compared with the DCQCN-like baseline, MatchRDMA improves inter-DC throughput by up to 20×.

Fig. 3(c) reports the destination-OTN runtime buffer occupancy. The DCQCN-like baseline allows excess inter-DC traffic to accumulate near the destination OTN before downstream constraints are fully reflected, resulting in larger buffer buildup. Fig. 3(d) shows the pause time ratio. Both the THEMIS-like baseline and MatchRDMA achieve lower pause overhead than the DCQCN-like and pseudo-ACK baselines, while MatchRDMA yields the lowest overall pause time ratio because it aligns source release with the destination-estimated rate budget. Compared with the DCQCN-like baseline, MatchRDMA reduces peak runtime buffer occupancy by up to 62.7% and the pause time ratio by up to 94.1%.

Fig. 3(e) shows the overall average flow completion time (FCT) under different message sizes in the mixed-traffic scenario. MatchRDMA outperforms the DCQCN-like baseline, reducing the average FCT by 31.5%–43.9%. The advantage becomes more significant for larger message sizes, because larger messages tend to produce more stable traffic patterns and are therefore easier to regulate through slot-level rate estimation.

Additional sweeps over traffic jitter and parallel-message concurrency show that MatchRDMA maintains the same performance trend and degrades more gracefully than the baselines.

**Conclusions**
We proposed MatchRDMA for geo-distributed LLM training over OTN. It addresses the challenges of long-haul RDMA transmission through source-OTN budget-gated pseudo-ACK generation and congestion-control proxying, inter-OTN rate-budget updates, and destination-OTN communication-aware slot-weighted rate estimation. MatchRDMA improves inter-DC throughput while reducing destination-OTN runtime buffer occupancy and pause time, and remains robust under distance variation, message-size scaling, and parallel-message concurrency.

**Acknowledgements**

This work was supported by the National Key R&D Program of China (No. 2024YFB2908303).